\documentclass[superscriptaddress, longbibliography, showkeys, prc]{revtex4-2}
\usepackage{eurosym}
\usepackage{amsmath,amsfonts,amssymb,amsthm,mathtools}
\usepackage{txfonts}
\usepackage{icomma}
\usepackage{fdsymbol}
\usepackage[dvipsnames]{xcolor}
\usepackage{graphicx}
\usepackage{dcolumn}
\usepackage{bm}
\usepackage{array,tabularx,tabulary,booktabs}
\usepackage{longtable}
\usepackage{multirow}
\usepackage[normalem]{ulem}

\graphicspath{{images/}}
\newcolumntype{M}[1]{>{\centering\arraybackslash}m{#1}}
\begin{document}

\title{Comment on the article "Detailed study of the astrophysical direct
capture reaction $^{6}$Li$(p,\gamma )^{7}$Be in a potential model approach" 
\textit{by E. M. Tursunov, S. A. Turakulov, and K. I. Tursunmakhatov}}

		\author{S. B. Dubovichenko}
		\affiliation{Fesenkov Astrophysical Institute$,$ 050020 Almaty$,$ Kazakhstan}
		\author{A.S. Tkachenko}
		\affiliation{Fesenkov Astrophysical Institute$,$ 050020 Almaty$,$ Kazakhstan}
		\author{R. Ya. Kezerashvili}
		\affiliation{New York City College of Technology$,$ City University of New York$,$ Brooklyn$,$ 11201 New York$,$ USA}
		\affiliation{Graduate School and University Center$,$ City University of New York$,$ 10016 New York$,$ USA}
		%\author{N.A. Burkova}
		%\affiliation{al-Farabi Kazakh National University$,$ 050040 Almaty$,$ Kazakhstan}

\begin{abstract}
We explicitly present the comparison of the results for the astrophysical $S-$factor and reaction rate for the  $^{6}$Li($p,\gamma $)$^{7}$Be capture
process at astrophysical energies, presented in Phys. Rev. Phys. Rev. C \textbf{105}, 065806 (2022) and  Phys. Rev. C \textbf{108}%
. 065801 (2023) obtained within the famework of potential models. We demonstrate that both potential model approaches describe successfully the astropgisical $S-$factor and reaction rate similtaneusly and reproduce the LUNA Collaboration [Phys.
Rev. C \textbf{102}, 052802(R) (2020)] results.
\end{abstract}
\preprint{APS/123-QED}
\maketitle

%\keywords{low and astrophysical energies, $p^{6}$Li system, thermonuclear reaction rate, potential cluster model}
%\preprint{APS/123-QED}

In the recent paper \cite{TTT2023} there is a statement: "Very recently, a
detailed study of the above $^{6}$Li($p,\gamma $)$^{7}$Be direct capture
process at astrophysical energies was performed within the potential model
[28]. Various versions of the potential model have been suggested; however,
none of them describe the astrophysical $S-$factor and the reaction rates
simultaneously. More precisely, the temperature dependence of the reaction
rates of the LUNA Collaboration [25] was not reproduced within that model.
Thus, the question of whether a potential model can simultaneously describe
the astrophysical $S-$factor and the reaction rates remains open."

Reference [28] is our article \cite{6Li2022}, and [25] is the LUNA
Collaboration \cite{Piatti2020}. This statement is written without making
any comparison between authors' results and results \cite{6Li2022} obtained
within a potential model approach. The statement motivated us to compare the
astrophysical $S-$factor and reaction rate obtained in Refs. \cite{6Li2022}
and \cite{TTT2023}. To shed light on the contradiction between this
statement and our results \cite{6Li2022}, we present a comparison of
calculations for astrophysical factor and reaction rate reported in \cite%
{6Li2022} and \cite{TTT2023} along with reported in literature experimental
data. 
\begin{figure}[b]
\centering
\includegraphics[width=8.9cm]{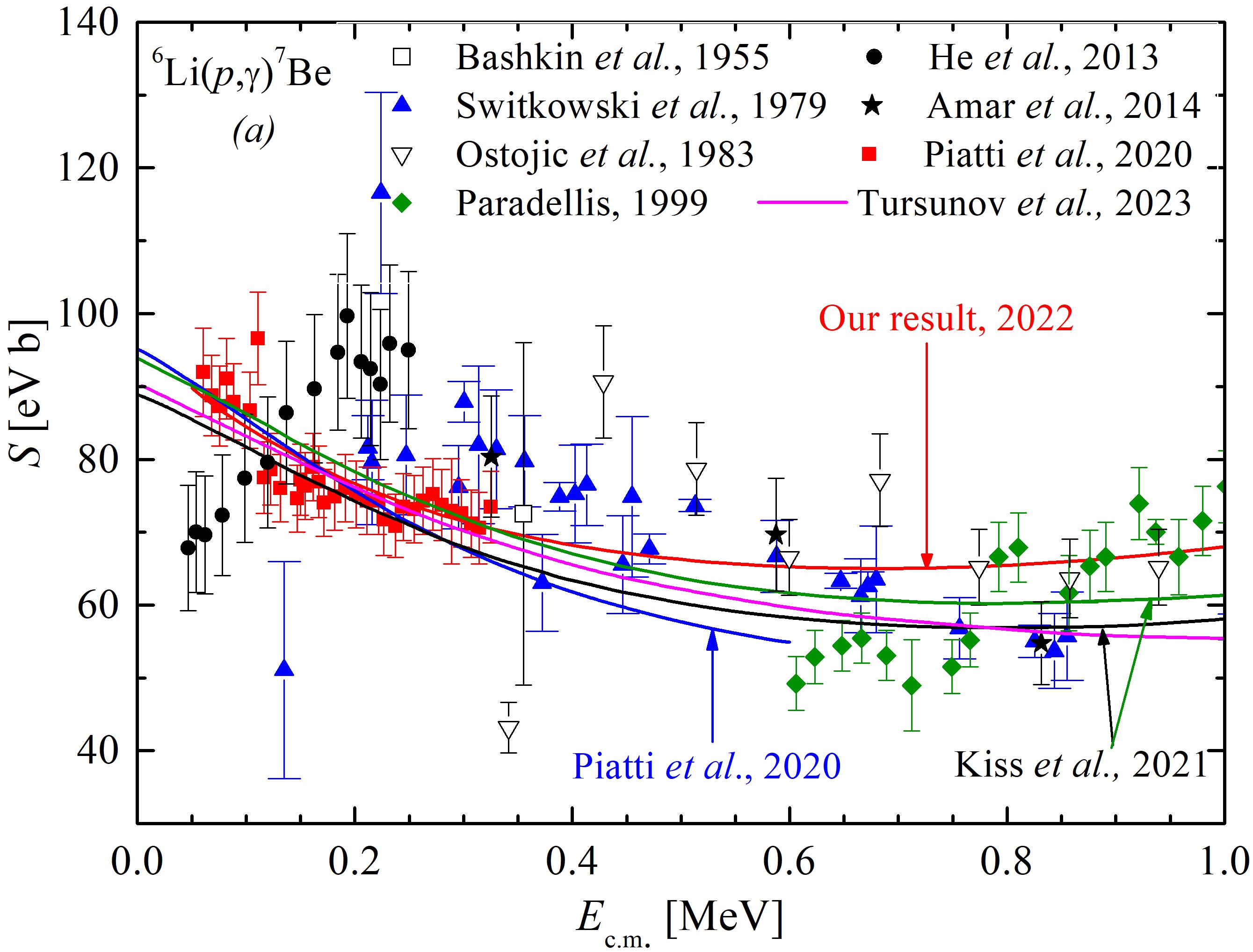}  \includegraphics[width=8.5cm]{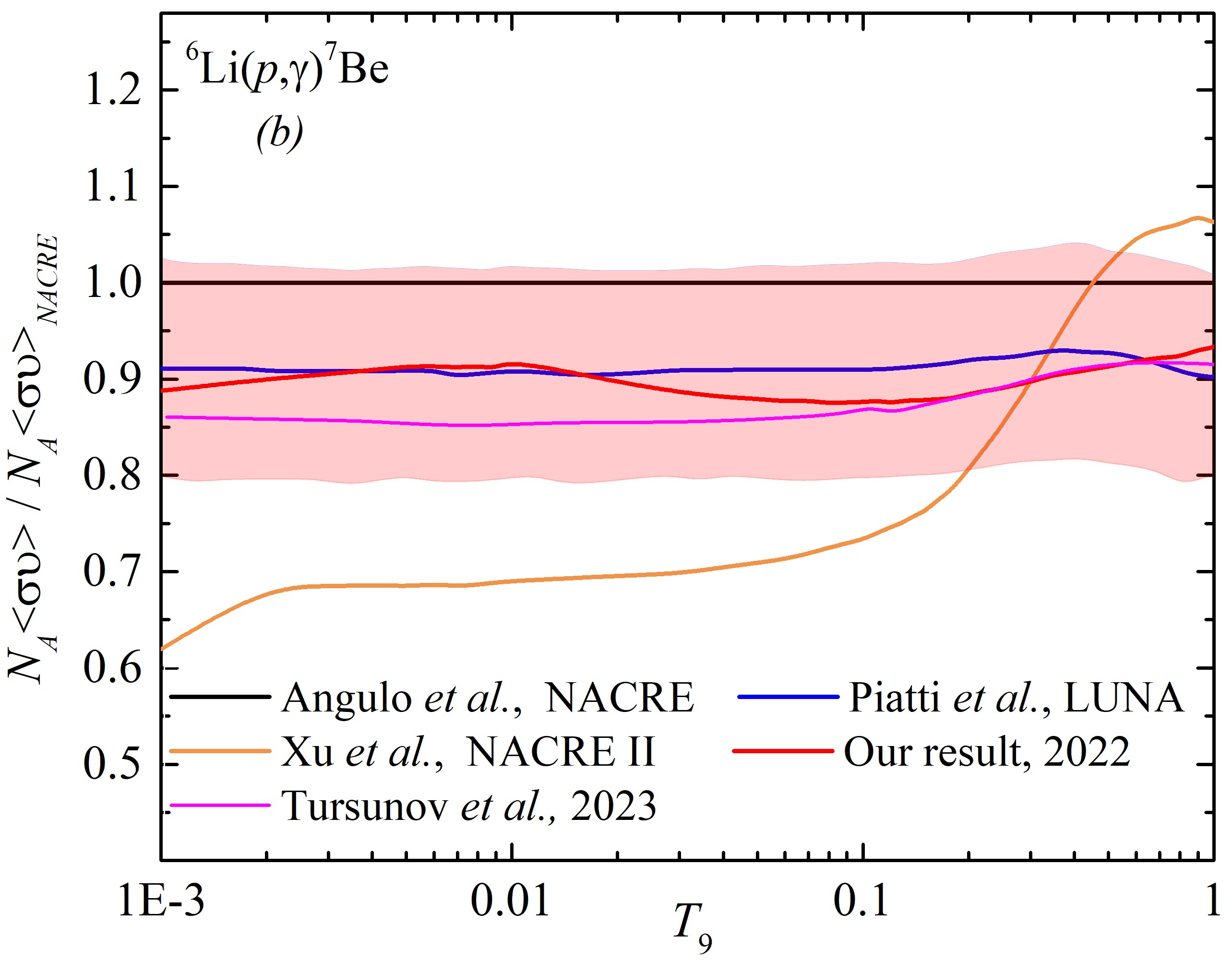}
\caption{(Color online) ($a$) Comparison of $^{6}$Li$(p,\protect%
\gamma )^{7}$Be reaction astrophysical $S-$factors. Experimental data are
from {\color{Blue}$\medblacktriangleup$} -- \protect\cite{Switkowski1979}, $
\medblackcircle$ -- \protect\cite{He2013}, {\ \color{Red}$\medblacksquare$}
-- \protect\cite{Piatti2020}, {\color{Green}$\medblackdiamond$} -- 
\protect\cite{Paradellis1999}, $\medsquare$ -- \protect\cite{Bashkin1955}, $
\medtriangledown$ -- \protect\cite{Ostojic1983}, $\medblackstar$ -- 
\protect\cite{Amar2014}. Results of calculations: red curve -- our work 
\protect\cite{6Li2022}; pink curve -- Ref. \cite{TTT2023}; blue curve -- Ref. \protect\cite{Piatti2020}; black
and green curves -- Ref. \protect\cite{Kiss2021}.  ($b$) Comparison of the
astrophysical reaction rates from 
\protect\cite{Xu2013,Piatti2020,6Li2022,TTT2023} in the range of 0.001 to 1 $T_{9}$, normalized to
the NACRE rate \protect\cite{Angulo1999}. The shaded area
represents the uncertainties from LUNA \protect\cite{Piatti2020} (pale red band).}
\label{Fig1}
\end{figure}

In Fig. 1$a$ we present the results of calculations for the astrophysical $S-$%
factor in the framework of different models and all reported experimental
data, including the most recent Luna Collaboration \cite{Piatti2020}
measurements. Our astrophysical $S-$factor is given with a solid red curve, 
and results \cite{TTT2023} are presented with the pink curve. The $R-$matrix fit of the
data from LUNA Collaboration \cite{Piatti2020} is represented with the solid
blue curve. A solid green curve was obtained by Kiss et al. \cite{Kiss2021}
using the weighted means of the ANCs from the analysis of the $^{6}$Li($^{3}$%
He,$d$)$^{7}$Be transfer reaction within the modified two-body potential
method (MTBPM). In addition, \cite{Kiss2021} contains the results for the $S-
$factor of the $^{6}$Li$(p,\gamma )^{7}$Be reaction calculated within the
MTBPM, using the values of ANCs obtained from the analysis of the
experimental astrophysical $S-$factors of the $^{6}$Li$(p,\gamma )^{7}$Be
reaction \cite{Piatti2020}. This result is given in Fig. 1$a$ with the solid
black curve. One can see that both \cite{6Li2022} and \cite{TTT2023}
potential model approaches provide results within the error bar of the LUNA
data \cite{Piatti2020}. 
There is a slight difference between
them, which is not surprising, since in \cite{TTT2023} slight modifications
of the potentials from \cite{6Li2022} were used. The uncertainty of experimental
data allows to select other options for the potentials to describe the LUNA
data \cite{Piatti2020}.
\iffalse
There is a slight difference between them,  
which is
not surprising, since in \cite{TTT2023} slight
modifications of potentials from \cite{6Li2022} were used. The uncertainty of experimental data allows to select other
options for the potentials to describe the LUNA data \cite{Piatti2020}. 
%The authors of \cite{TTT2023} did not use all the available experimental data,
which led to a slightly different form of the $S-$factor at energies above
0.4 MeV than our result \cite{6Li2022} and results of the $R-$matrix fit of
the data  \cite{Piatti2020} and prediction \cite{Kiss2021}. 

\fi
Thus, the potential
model \cite{6Li2022} reproduces LUNA Collaboration \cite{Piatti2020} data
for the astrophysical $S-$factor.% and beyond.

Next, we compare the reaction rates from Refs. \cite{6Li2022} and \cite%
{TTT2023}. A comparison of the reaction rates from\ \cite{6Li2022} and \cite%
{TTT2023} with the experimental data of the LUNA Collaboration \cite%
{Piatti2020} and the results of the NACRE II compilation \cite{Xu2013},
normalized to the NACRE rate \cite{Angulo1999} in the temperature range of $%
T_{9}=0.001-1$ is shown in Fig. 1$b$. The shaded area represent the
uncertainties from the LUNA Collaboration \cite{Piatti2020}. The deviation
between the adopted reaction rate obtained in \cite{Piatti2020} and our
calculation \cite{6Li2022} in the range of $T_{9}=0.001-1$ does not exceed
5\%, while the deviation of the results \cite{TTT2023} in a temperature range
of $T_{9}=0.001-0.015$ is larger. The analytical approximations of the
reaction rates reported in \cite{Piatti2020}, \cite{6Li2022}, and \cite%
{TTT2023} are different, but the corresponding reaction rates are within the
uncertainties given by the LUNA Collaboration \cite{Piatti2020}.

Thus, both \cite{6Li2022} and \cite{TTT2023} single-channel potential model
approaches reproduce not only the absolute values of the reaction rates of Ref. \cite{Piatti2020} but also the temperature dependence of the reaction rates. Consequently, both \cite%
{6Li2022} and \cite{TTT2023} potential model approaches can describe simultaneously 
the $S-$factor and reaction rate reported by the LUNA Collaboration \cite{Piatti2020} for the $^{6}$Li($p,\gamma $)$^{7}$Be reaction. Therefore, it's not clear how the results of \cite{TTT2023} differ from ours \cite{6Li2022} and what is the novelty of \cite{TTT2023}. In addition, it is important to mention that the advantage of the potential model approach in \cite{6Li2022} relates to a realistic description of the $p^6$Li interaction in scattering and bound-state channels and it does not need the use of any additional fitting parameters for reproducing experimental data in contrast to \cite{TTT2023}.

\end{document}